\documentclass[%
 reprint,
 amsmath,amssymb,
 aps,
pra,
]{revtex4-2}

\usepackage{graphicx}
\usepackage{dcolumn}
\usepackage{bm}

\usepackage{xcolor}
\usepackage[switch]{lineno}

\begin{document}

\title{Phase statistics of a single qubit emission as a direct probe of its coherence}

\author{A.~Sultanov}
\affiliation{Leibniz Institute of Photonic Technology, D-07745 Jena, Germany}
\email{aidar.sultanov@leibniz-ipht.de}

\author{E.~Mutsenik}
\affiliation{Leibniz Institute of Photonic Technology, D-07745 Jena, Germany}
\author{L.~Kaczmarek}
\affiliation{Leibniz Institute of Photonic Technology, D-07745 Jena, Germany}
\author{M.~Schmelz}
\affiliation{Leibniz Institute of Photonic Technology, D-07745 Jena, Germany}

\author{G.~Oelsner}
\affiliation{Leibniz Institute of Photonic Technology, D-07745 Jena, Germany}

\author{R.~Stolz}
\affiliation{Leibniz Institute of Photonic Technology, D-07745 Jena, Germany}

\author{E.~Il'ichev}
\affiliation{Leibniz Institute of Photonic Technology, D-07745 Jena, Germany}

\begin{abstract}

The emission of photon from an individual atom encodes the phase of its initialized quantum state. Using single-shot heterodyne detection, we measure the phase distribution of the emission from a superconducting transmon qubit in an open waveguide configuration and track its evolution over time.
We demonstrate that the presence of a quantum superposition is encoded in the phase statistics of the emission and remains resolvable despite a high noise level.
These phase statistics serve as a quantitative probe of the qubit coherence. The decay of the emission envelope with increasing integration time reveals the energy relaxation rate of the emitted wavepacket, while phase distribution broadening  tracks pure dephasing processes. We thereby establish a direct link between the decoherence dynamics of an open quantum system and the statistical properties of its radiated field.

\end{abstract}

\maketitle

Quantum coherence in two-level systems is manifested in a well-defined relative phase between its basis states. In experimental practice, coherence is typically characterized through measurements performed directly on the emitter, such as quantum state tomography~\cite{Bianchetti2010,Liu2004} or Ramsey interferometry~\cite{Bylander2011,Yoshihara2006}. 
Alternatively, according to the principles of light–matter interaction this coherence is imprinted onto the emitted radiation. This mapping has been exploited to reconstruct the emitter's quantum state and its trajectories from measurements of the outgoing field, for instance using continuous homodyne or heterodyne detection in cavity quantum electrodynamics settings~\cite{Ficheux2018,Campagne-Ibarcq2016,Naghiloo2016}.

In open quantum systems, such as a qubit coupled to an open waveguide, the radiated field is the primary and often the only experimentally accessible observable~\cite{Abdumalikov2011,Sheremet2023}. Due to non-classical correlations in observables and the probabilistic nature of quantum physics, the field's statistical properties are crucial for inferring the emitter's dynamics. For weak, stochastically emitted signals dominated by additive noise, a key challenge is to identify which measurable  statistical features of the measured field provide the most robust and direct characterization  of the emitter's coherence. We propose that the phase statistic of the emitted field,  quantified by its concentration or first-order circular moment, represents  such a characteristic. We investigate whether this moment can serve as a direct, noise-resilient measure of the coherent component of the emission and whether its temporal evolution can track the emitter's coherence dynamics.

To explicitly establish the connection between the emitter's superposition state and the phase concentration of emission, we first consider a simplified ideal case. The fundamental link between the superposition and the phase of its emission is captured by a minimal model of light–matter interaction. We consider a two-level system coupled to a single electromagnetic mode.  Within the rotating-wave approximation, the corresponding interaction Hamiltonian reads:

\begin{equation}
H_{\mathrm{int}}=\hbar \lambda\left(a^{\dagger}\sigma_-+a\sigma_+\right),
\end{equation}
where $\sigma_-$ ($\sigma_+$) are the lowering (raising) operators of the qubit, and $a^\dagger$ ($a$) are the creation (annihilation) operators of the field mode, and $\lambda$ is the coupling strength. We assume that the initial state of the two-level quantum system is general:
\begin{equation}
\label{eq:superpos_general}
\lvert\Psi(0)\rangle=
\left(\cos\left(\frac{\theta}{2}\right)  \lvert g\rangle + \sin\left(\frac{\theta}{2}\right) e^{i\varphi}\lvert e\rangle\right)\otimes\lvert 0\rangle,
\end{equation}
where $\lvert e\rangle$ and  $\lvert g\rangle$ are ground and excited states of the system, $\theta$ and $\varphi$ are angles of the state vector on the Bloch sphere,  $\lvert 0\rangle$ is the vacuum state of the field, and $i$ is the imaginary unit. The time evolution of $\lvert\Psi(0)\rangle$ is then described by a formal solution of the time-dependent Schr\"odinger equation, $\lvert\Psi(t)\rangle=e^{-iH_{\mathrm{int}}t/\hbar}\lvert\Psi(0)\rangle$:
\begin{equation}
\begin{aligned}
\lvert\Psi(t)\rangle &= 
\cos\left(\frac{\theta}{2}\right)\lvert g,0\rangle
 + \sin\left(\frac{\theta}{2}\right) e^{i\varphi}
\Bigl[
\cos\left(\lambda t\right)\lvert e,0\rangle \\
&\quad - i \sin\left(\lambda t\right)\lvert g,1\rangle
\Bigr].
\end{aligned}
\end{equation}

Thus, the expectation value of the emitted field is:

\begin{equation}
\begin{aligned}
    \label{eq:emission}
\langle a\rangle &= \langle \Psi(t)  \lvert a\lvert\Psi(t)\rangle=-\frac{i}{2}\sin(\theta)\sin(\lambda t) e^{i\varphi}\equiv \\
&\quad \langle I\rangle+i \langle Q\rangle
\end{aligned}
\end{equation}
where $I$ and $Q$ are the field quadratures. After amplification and detection, the field amplitude $(I^2+Q^2)^{1/2}$ and phase  $\varphi = \arctan(-I/Q)$ can be obtained. This minimal model captures the essential phase mapping, a more complete description of emission into an open waveguide requires a treatment of the full electromagnetic continuum \cite{Clerk2010, Lalumiere2013,Greenberg2024,Greenberg2025}.

\begin{figure*}
    \centering
    \includegraphics[width=1\textwidth]{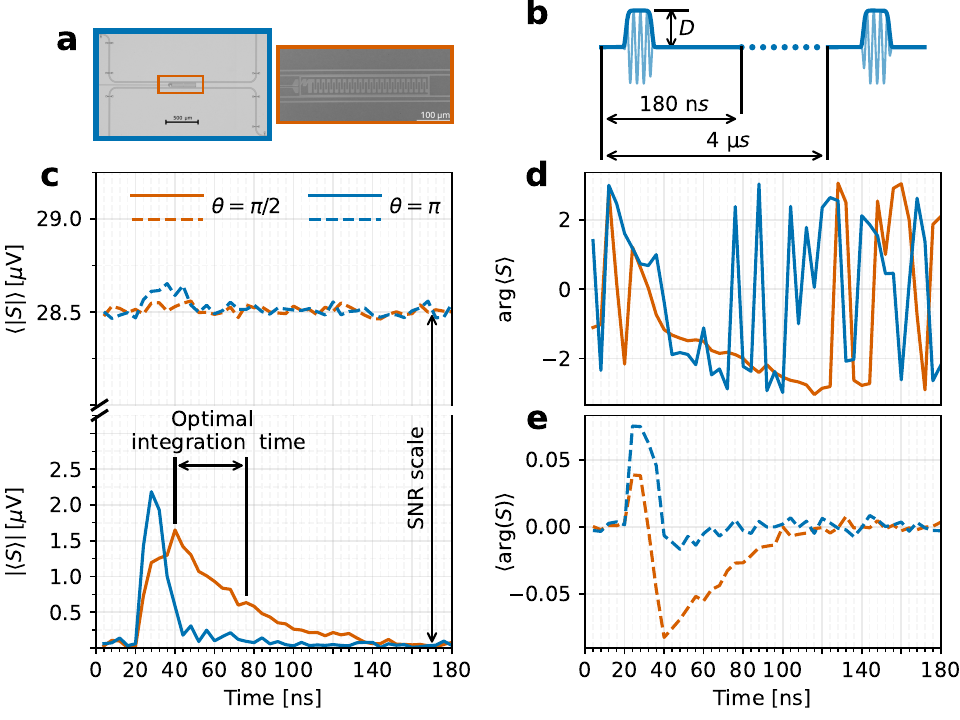}
\caption{
Time-resolved emission of a qubit under controlled drive pulses. 
\textbf{a}, The optical image of the device showing two waveguides coupled to a transmon qubit with a zoomed image on the transmon region obtained with scanning electron microscope. 
\textbf{b}, Example pulse sequence used to prepare the qubit with pulse amplitude $D$, duration of 180~ns per repetition, and 4~$\mu$s spacing between repetitions. 
\textbf{c}, The signal amplitudes: $\langle |S| \rangle$ (dashed lines) represents the shot-averaged magnitude, dominated by background noise, while $|\langle S \rangle|$ (solid lines) reveals coherent emission emerging from complex-plane averaging, canceling incoherent background; the vertical double-headed arrow indicates the signal-to-noise ratio (SNR) scale, and the horizontal arrow marks the optimal time window for analysis. 
\textbf{d}, Phase of the shot-averaged signal, $\arg \langle S \rangle$, exhibiting minimal information and a slight drift, demonstrating residual readout-related instabilities. 
\textbf{e}, Mean phase, $\langle \arg(S) \rangle$, showing the main result: coherent emission with phase evolution dependent on qubit preparation angle, illustrating the temporal signature of the emitted radiation. 
}
\label{fig:emission_time_domain}
\end{figure*}

For an experimental study we fabricated a system consisting of two open waveguides coupled to a superconducting transmon qubit. Here, one waveguide used as a drive line for the qubit preparation, while the second one is used to collect the emission. For each  $n$-th experimental realization, we record the complex field $S_n(t) = I_n(t) + i Q_n(t)$ (see Fig.~\ref{fig:emission_time_domain}a,b and Methods for detailed information about the experiment). Panels c-e of Fig.~\ref{fig:emission_time_domain}  present the field parameters for two qubit states, $\theta = \pi$ and $\theta = \pi/2$, measured over $N=5 \times 10^5$ repetitions. The mean magnitude, $\langle |S| \rangle$, remains at the noise level for both states, whereas the phase-sensitive averaging, $|\langle S \rangle|$, reproduces the applied pulse envelope and, for $\theta = \pi/2$, exhibits a long decaying tail corresponding to coherent emission (see Fig.\ref{fig:emission_time_domain}c). The phase of the mean, $\arg \langle S \rangle$ in Fig.\ref{fig:emission_time_domain}d, shows a slow drift over the measurement window, consistent with residual readout-related phase instabilities. In contrast, the mean phase, $\langle \arg(S) \rangle$ shown in Fig.\ref{fig:emission_time_domain}e, has the expected temporal behavior of the emission: for $\theta = \pi$ it is nonzero only during the drive pulse, while for $\theta = \pi/2$ it decays over the characteristic decoherence time $\approx 40$~ns. These examples illustrate typical features of the measured traces and the signal-to-noise ratio (SNR) scale, providing context for the subsequent statistical analysis. The time-domain data show that the emitted radiation contains a coherent component whose phase depends on the preparation angle $\theta$.

Since the signal obtained in single shot is dominated by measurement noise (see SNR  scale in Fig.~\ref{fig:emission_time_domain}c), averaged traces alone do not fully capture the underlying phase relations.  To access the coherent contribution at the single-shot level, we analyze the statistical distribution of the measured signal over $N$ realizations.

\begin{figure}
    \centering
    \includegraphics[width=1\columnwidth]{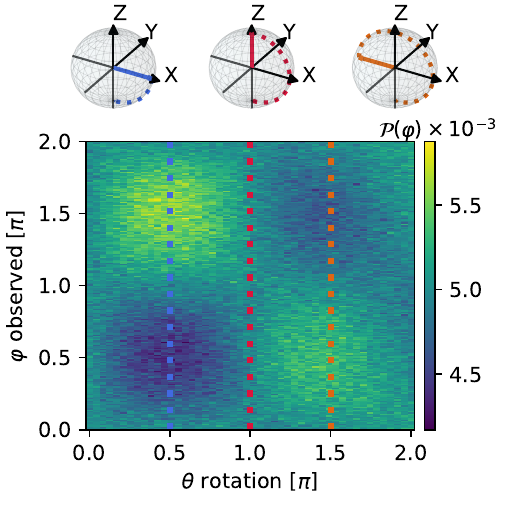}
\caption{
Phase probability density of the emitted microwave field for different qubit preparation angles $\theta$. The probability density $\mathcal{P}(\varphi)$ is constructed from repeated measurements of the field phase $\varphi \equiv \arg(S)$, obtained by integrating the emission over the optimal 36~ns time window indicated in Fig.~\ref{fig:emission_time_domain}c. The heatmap shows $\mathcal{P}(\varphi)$, with vertical lines marking the three preparation angles $\theta=\pi/2$ (blue, oriented along $+x$), $\pi$ (red, $+z$), and $3\pi/2$ (orange, $-x$), corresponding to the Bloch vectors depicted on the adjacent Bloch spheres. For $\theta=\pi/2$ and $\theta=3\pi/2$, the distributions exhibit pronounced peaks separated by $\pi$, reflecting the sign change of the coherent emission amplitude. For $\theta=\pi$, the distribution is nearly uniform, consistent with the absence of a well-defined emission phase. These results show that the phase statistics of the emitted field faithfully reflect the prepared qubit superposition.}
    \label{fig:pdf_rotations}
\end{figure}

For each realization, the emitted field is integrated over an optimally chosen time window $T=36$~ns, yielding a set of complex values $\tilde{I}_n + i \tilde{Q}_n$ (see Methods).
 The corresponding phase samples
\[
\varphi_n = \arg(\tilde{I}_n + i\tilde{Q}_n)
\]
define the phase probability density $\mathcal{P}(\varphi)$.
While amplitude statistics are strongly masked by noise, the phase distribution reveals a clear structure, see Fig.~\ref{fig:pdf_rotations}. Coherent emission produces a peaked phase distribution, whereas purely incoherent noise yields a uniform distribution. For $\theta=\pi/2$ (blue), the phase probability density function exhibits a pronounced peak around $\varphi=3\pi/2$, while for $\theta=3\pi/2$ (orange) the peak is shifted by $\pi$, appearing near $\varphi=\pi/2$, fully consistent with expectations. This phase shift directly reflects the sign change of $\sin\theta$ in Eq.~(\ref{eq:emission}). For $\theta=\pi$, the coherent emission vanishes and the phase distribution becomes uniform, indicating a fully random phase. Although the coherent emission exhibits a well-defined mean field, Eq.~(\ref{eq:emission}) assumes a sharp dependence on $\theta$, whereas the experimental data exhibit a finite phase spread.

\begin{figure}
    \centering
    \includegraphics[width=1\columnwidth]{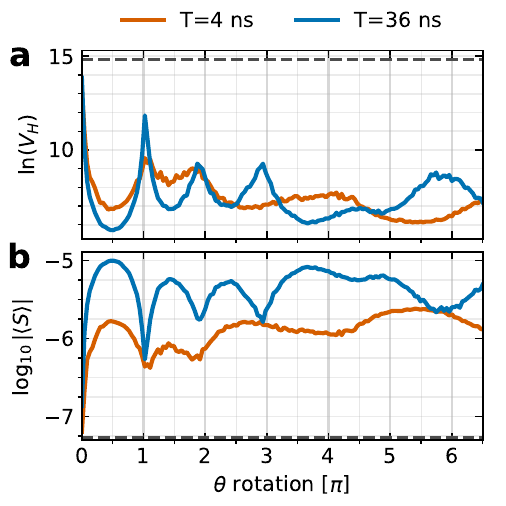}
\caption{
Dependence of qubit emission amplitude and phase uncertainty on the qubit preparation angle $\theta$ and integration time window $T$. 
\textbf{a}, Holevo variance $V_H$ of the emitted field, quantifying phase uncertainty, as a function of $\theta$. $V_H$ is calculated from repeated measurements of the emission phase $\varphi \equiv \arg(S)$, integrated over the indicated temporal windows (solid curves).  Maxima of $V_H$ occur at integer multiples of $\pi$, reflecting maximal phase uncertainty when the emitted field lacks a well-defined phase. 
\textbf{b}, Mean emission amplitude corresponding to the same dataset. Minima appear at odd multiples of $\pi$, indicating incoherent emission, and at even multiples of $\pi$, corresponding to the qubit prepared in the ground state. Dashed horizontal lines in both panels denote reference levels measured prior to qubit manipulation.  Different solid curves correspond to distinct integration times, highlighting the influence of temporal window choice. The anticorrelated behavior of amplitude and Holevo variance reflects the inverse relationship between coherent emission strength and phase uncertainty, while deviations at larger $\theta$ indicate intrinsic qubit nonlinearities and non-stationary processes within the integration window.}
    \label{fig:Correlation_VH_mean_field_amplitude}
\end{figure}

This spread arises from noise that broadens the measured phase distribution. Since the phase is a circular variable, we use the Holevo variance \cite{Costa2021,Berry2009} to quantify this broadening:
\begin{equation}
V_H = R^{-2} - 1,
\end{equation}
where $R = \left| \frac{1}{N} \sum_{n=1}^{N} e^{i \varphi_n} \right|$ denotes the mean resultant length which is a characteristic of the sharpness of the phase distribution. Here, ${\varphi_n}$ denotes the ensemble of $N$ phase measurements obtained from single-shot emission events. For an ideal coherent emission, $R=1$, yielding a vanishing Holevo variance. The dependence of $V_H$ on the rotation angle $\theta$ is shown in Fig.~\ref{fig:Correlation_VH_mean_field_amplitude}a for two different integration times. In the ideal case of a linear two-level system driven by resonant pulses, one would expect $V_H$ to oscillate with $\theta$, reaching maxima at odd multiples of $\pi$, where the emission is incoherent. Correspondingly, the mean emission amplitude (Fig.~\ref{fig:Correlation_VH_mean_field_amplitude}b) shows the opposite behavior, with minima at odd multiples of $\pi$. Two key observations emerge from the data: first, at larger rotation angles, corresponding to higher drive pulse amplitudes, the nonlinear behavior becomes more pronounced. This behavior may originate from intrinsic qubit nonlinearities, such as higher excited states and dressing effects, as well as from the up-conversion chain. Second, the observed dependence changes with increasing integration time, indicating the presence of non-stationary processes within the integration window at large drive amplitudes. 

To focus exclusively on stationary and approximately linear processes, we therefore restrict further analysis to datasets acquired at $\theta=\pi/2$. 
The qubit emission from the initial state Eq.~(\ref{eq:superpos_general}) is probabilistic in nature. Moreover, the qubit can emit into two different waveguides with the probabilities determined by the waveguide-qubit coupling strengths. These conditions imply that the probability $p$ of detecting an output signal at the selected port is less than unity.
By assuming that the inevitable noise is white, additive and follows complex normal distribution with zero mean and variance $\sigma_N/2$ per quadrature, denoted as $\mathcal{CN}(0,\sigma_N/2)$,  
a single-shot measured signal can be expressed as follows:

\begin{equation}
\begin{aligned}
    \tilde{I}_n + i\tilde{Q}_n= S_{qb} X+\mathcal{CN}(0,\sigma_N/2),\\
\end{aligned}
\end{equation}
where $S_{qb}=\tilde{I}_{qb} + i\tilde{Q}_{qb}$ is the value of the field emitted by the qubit and $X$ is a binary random variable, describing stochastic single-shot emission events with $P(X=1)=p$. For $p=1$, this model reduces to the conventional description of a deterministic signal measured in the presence of Gaussian noise. However, the proposed formulation explicitly takes into account the probabilistic nature of the detected emission.

The statistics of the single-shot measurement determine the properties of the signal averaged over $M$ independent repetitions.
Let $k$ denote the number of realizations (out of $M$) in which emission is detected such that $k=\Sigma_{i=1}^{M} X_i.$ Since each repetition $X_i$ constitutes an independent Bernoulli trial with success probability $p$, the number of emission events $k$ follows a binomial distribution,
\begin{equation}
P(k) = \binom{M}{k} p^k (1-p)^{M-k}.
\end{equation}
The complex signal averaged over $M$ repetitions, conditioned on $k$, is
\begin{equation}
\langle\tilde{I}_n + i\tilde{Q}_n\rangle_M = \frac{k}{M} \left(\tilde{I}_{qb} + i\tilde{Q}_{qb}\right) + \bar{N},
\end{equation}
where $\bar{N}$ is Gaussian noise with variance reduced by a factor $M$ due to averaging. The corresponding effective SNR is therefore
\begin{equation}
\rho_k = \frac{k \eta}{ \sqrt{M}},
\end{equation}
where $\eta=\frac{|S_{qb}|}{\sigma_N}$. For a deterministic complex signal measured in circular Gaussian noise the effective SNR reads as: $\rho =  \eta \sqrt{M}$. This explicitly shows that probabilistic detection degrades the SNR since $k<M$.

For the fully deterministic signal the mean resultant length admits the analytical expression \cite{Presnell2008}
\begin{equation}
R_{det}(\rho)
=
\frac{\sqrt{\pi}}{2}
\rho
e^{-\rho^2/2}
\left[
I_0\left(\frac{\rho^2}{2}\right) + I_1\left(\frac{\rho^2}{2}\right)
\right],
\end{equation}
where $I_0$ and $I_1$ are modified Bessel functions.
Averaging over the binomial distribution of $k$ yields the predicted mean resultant length
\begin{equation}
\label{eq:Sharpness}
R(M,p)
=
\sum_{k=0}^{M}
P(k)
\, R_{det}(\rho_k).
\end{equation}
Note that for negligible Gaussian noise the effective SNR approaches infinity $\lim_{\eta\rightarrow \infty} \rho_k=\infty$ and, consequently, $\lim_{\eta\rightarrow \infty} R_{det}(\rho_k)=1$. Therefore, a signal can be detected in a single-shot measurement~$\displaystyle{M=1}$. In this case, Eq.~(\ref{eq:Sharpness}) simplifies to $\lim_{\eta\rightarrow \infty}R(M=1,p)=p $. Thus, in the high SNR limit, the mean resultant length $R$ directly approaches the probability of the stochastic detection $p$.

\begin{figure*}[!htbp]
    \centering
    \includegraphics[width=1\textwidth]{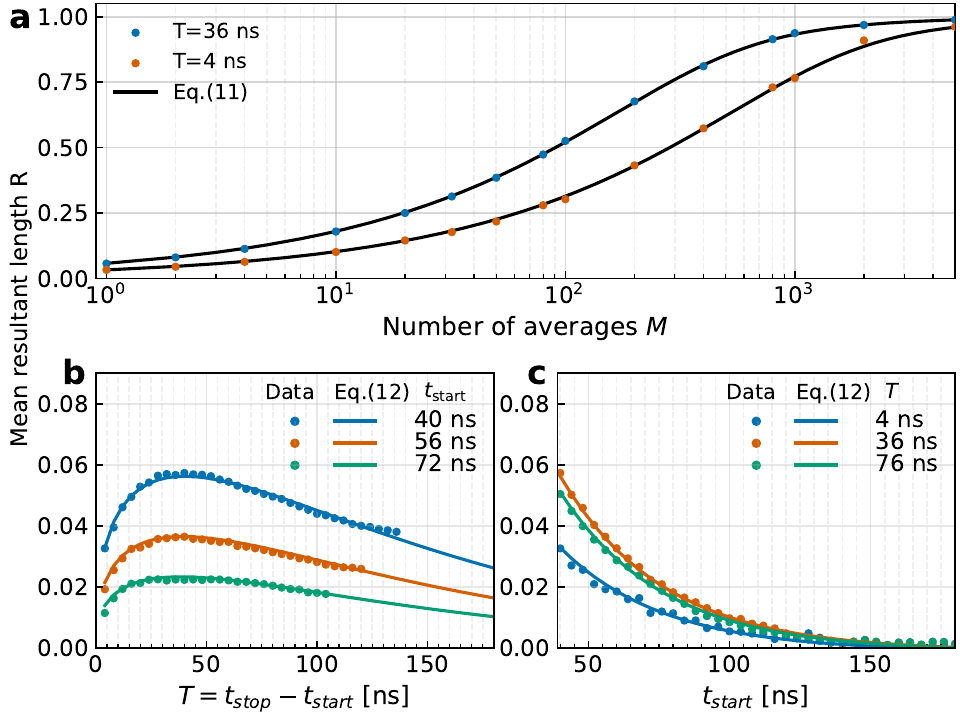}
\caption{
Mean resultant length $R$ of the emitted qubit radiation, quantifying the sharpness of the phase distribution. 
\textbf{a}, $R$ as a function of the number of averages $M$ for integration times of 4~ns and 36~ns, demonstrating how averaging over repeated measurements increases phase concentration. 
\textbf{b}, $R$ versus integration time $T$ for different observation start times $t_{\rm start}$. Solid lines are fits using the phenomenological model of Eq.~(\ref{eq:Sharpness_1}), capturing the combined effects of wavepacket decay and phase decoherence. 
\textbf{c}, $R$ as a function of $t_{\rm start}$ for several fixed integration times, showing the decay of phase coherence along the wavepacket envelope. 
These panels together illustrate how the detected phase sharpness depends on averaging, observation window, and temporal position along the emission, revealing both the wavepacket envelope and the influence of finite phase coherence on the measured signal.
}
    \label{fig:VH_scaling}
\end{figure*}

In practice, we process the complete data sent of  $N=5\times10^5$ single-shot realizations by dividing it into non-overlapping groups of $M$-repetitions. Within each group, the complex signals are averaged before phase extraction. Since all single-shot measurements are statistically independent and the ensemble is sufficiently large, this procedure is mathematically equivalent to directly constructing an ensemble of signals averaged over $M$ repetitions. The resulting reduced ensemble of $N/M$ phase samples faithfully represents the distribution of $M$-shot averages without bias. The mean resultant length $R$ is then evaluated from this reduced ensemble. Repeating this procedure for different $M$ yields the experimental dependence $R(M,p)$, which is fitted using Eq.~(\ref{eq:Sharpness}) with $\eta$ and $p$ as free parameters and shown in Fig.~\ref{fig:VH_scaling}a. 

However, despite its apparent two-parameter form, Eq.~(\ref{eq:Sharpness}) depends only on the product $p\eta$. This reflects the internal structure of the model: probabilistic emission effectively rescales the coherent amplitude by a factor $p$, so that $R(M)$ is determined by the effective SNR $\eta p\sqrt{M}$. Fitting the measured $R(M)$ values for two integration windows yields $p\eta \approx 0.038$ for $T=4$~ns and $p\eta \approx 0.066$ for $T=36$~ns. If the emitted field was perfectly coherent throughout the integration window, one would expect $p\eta$ to scale as $\sqrt{T}$, predicting a threefold increase from 4~ns to 36~ns. The observed increase is significantly smaller, indicating a partial loss of signal coherence during integration. This deviation naturally demonstrates the presence of decoherence or other time-dependent processes that limit the effective accumulation of phase-coherent amplitude. 

To quantify this influence, we constructed a phenomenological model based on the following considerations: 
\begin{itemize}
    \item After the qubit enters the free evolution, it emits a wavepacket into the waveguide. The amplitude envelope of this wavepacket is set by the qubit’s energy relaxation rate. We have recently shown that for the current geometry, the qubit relaxation rate is determined predominantly by its coupling to the waveguide \cite{Mutsenik2025}. Since the emitted power follows the excited state population, it decays as $e^{-t/T_1}$ with $T_1$ the energy relaxation time. The field amplitude is proportional to the square root of the emitted power, so that the wavepacket envelope is expected to decay as $e^{-t/(2T_1)}$.
    \item During the emission process, the qubit undergoes dephasing processes that randomly distort the phase of the emitted field. This reduces the signal's phase concentration over time. Thus, the sharpness $R$ decays at a rate determined by the pure dephasing rate of the qubit.
    \item Integrating the noisy signal over a finite time window reduces the influence of the external noise. This introduces a growth of the measured sharpness of the phase distribution with the length of the integration window, following a sublinear scaling.
\end{itemize}

Motivated by these considerations, we introduce the following phenomenological form for the mean resultant length as a function of the observation start time $t_{start}$ and the integration time $T$:
\begin{equation}
\label{eq:Sharpness_1}
    R\left(t_{start},T\right)=A e^{-\frac{t_{start}}{\tau_{1}}}T^\beta e^{-\frac{T}{\tau_{2}}}+C
\end{equation}
where $\tau_1$ captures the wavepacket amplitude decay, $\tau_2$ captures decay of phase coherence, $\beta$ characterizes the noise's color or memory ($\beta=0.5$ for white noise) and $A,C$ are scaling and offsets, respectively. 
To verify that the $t_{\mathrm{start}}$ and $T$ dependencies can be treated independently, we performed a singular value decomposition of the measured two-dimensional $R(t_{\mathrm{start}},T)$ surface. The leading singular value captures the majority of the variance, justifying the factorization ansatz. The complete two-dimensional dependence $R(t_{start},T)$ was fitted simultaneously using a single global set of parameters $(A,\tau_1,\beta,\tau_2,C)$.  The fitted dependencies are shown in Fig.~\ref{fig:VH_scaling}b, c for $M=1$, giving $\tau_1\approx37$~ns, $\tau_2\approx105$~ns and $\beta\approx0.37$. These values are consistent with previously reported characteristic times \cite{Mutsenik2025}.  While $\beta=0.5$ corresponds to white, uncorrelated noise, the observed sub–$\sqrt{T}$ scaling indicates temporally correlated phase fluctuations. This interpretation is consistent with the slow evolution of the averaged phase $\text{arg}\langle S\rangle$ shown in Fig.~\ref{fig:emission_time_domain}d. Similar low-frequency noise processes are commonly observed in superconducting qubits \cite{Yan2012,Schlor2019,Anton2012}. Interestingly, the functional form of Eq.~(\ref{eq:Sharpness_1}) mirrors the non-diagonal elements of density matrix for two-level system, which reads \cite{NielsenChuang2010}:
\begin{equation*}
  \rho_{01}(t)=\rho_{01}(0)e^{-\frac{t}{2T_1}}e^{-\frac{t}{T_{\varphi}}}
\end{equation*}
where $T_1,T_\varphi$ stand for energy relaxation and pure dephasing times. This highlights that the mean resultant length $R$ naturally tracks the evolution of the coherence term of the density matrix. In this sense, $R$ constitutes a physically meaningful and experimentally accessible probe of phase coherence, reflecting both relaxation and dephasing dynamics at the level of ensemble phase statistics.

To conclude, we have demonstrated that the phase of the radiation emitted by a superconducting qubit directly reflects the initialized qubit's superposition. By constructing single-shot measured phase histograms and quantifying phase coherence via the Holevo variance, we showed that the phase distribution encodes the qubit’s superposition state, with sharpness correlated to the mean emission amplitude. Analysis of the dependence on the number of repetitions and on time reveals the stochastic nature of single-shot emission, as well as the temporal evolution of phase coherence, including the effects of dephasing and low-frequency correlated fluctuations. 

Furthermore, these results provide a direct, time-resolved window into the quantum coherence dynamics of a single qubit, establishing a quantitative link between single-shot phase statistics and underlying decoherence processes. The observed characteristics follow from the dynamics of an open two-level system and do not rely on specific microscopic details of the transmon-based implementation.

\begin{acknowledgments}
This work was partially supported by the German Federal Ministry of Education and Research under Grant Nos. 13N16152/QSolid and the European Innovation Council’s Pathfinder Open programme under grant agreement number 101129663/QRC-4-ESP. A.S. is grateful to Ya. Greenberg for fruitful discussions. 
\end{acknowledgments}
\section*{Author Contributions}
A.S: Conceptualization (equal), Data curation (lead), Formal analysis (lead), Investigation (equal), Methodology (equal), Validation (equal), Visualization (equal), Writing-original draft (equal), Writing-review and editing(equal); E.M: Conceptualization (equal), Data curation (supporting), Formal analysis (supporting), Investigation (equal), Methodology (supporting), Visualization (equal), Writing-review and editing (equal); L.K.: Resources (equal), Visualization (supporting), Writing-review and editing (equal); M.S.: Resources (equal),Writing review and editing (equal); G.O.: Funding acquisition (equal), Project administration (equal), Writing review and editing (equal); R.S.: Funding acquisition (equal), Project administration (equal), Writing-review and editing (equal); E.I: Conceptualization (equal), Formal analysis (supporting), Methodology (equal), Supervision, Validation (equal), Writing-original draft (equal), Writing-review and editing (equal).

\bibliography{nat_refs}

\clearpage
\section*{Methods}

To detect the emission from a two-level system, we fabricated a structure called a quantum microwave routing basic cell \cite{Mutsenik2025}. 
Here, the qubit is placed between two open microwave coplanar waveguides with finite coupling to both. This structure was fabricated from aluminum on a silicon substrate \cite{Schmelz2024}. Two identical waveguides with a central linewidth of 10~$\mu$m and gaps of 5~$\mu$m to ground planes were designed to ensure 50~$\Omega$ impedance matching. A transmon-type qubit \cite{Koch2007} was fabricated using the so-called Manhattan technology. To suppress parasitic resonances, aluminum air bridges were used \cite{Kaczmarek2025}. One waveguide is used for qubit initialization: a coaxial attenuated line is connected to one of its ports to deliver the drive, while the second port is connected to a 50~ohm‑matched amplified line that remains unused during initialization.
The other waveguide collects the emitted power and directs it to the detector via one of its ports; the second port is terminated with a 50~ohm attenuated line (see Fig.~\ref{fig:emission_time_domain}a). The qubit fundamental  transition frequency is $\omega_{01}=6.2503$~GHz, the relaxation time $T_1$=20.5~ns, driven Rabi oscillation decay time $T_R=25$~ns.  

The sample is installed in a dilution refrigerator with a base temperature of 10~mK. The signals from room temperature are delivered to the sample via --110~dB attenuated coaxial lines. A conventional cryogenic semiconductor amplifier with a noise temperature of about 2~K was used as a first stage of amplification. We employ a heterodyne scheme for both up- and down-conversion, implemented using an integrated quantum control platform (Quantum Machines OPX+ and Octave) to probe the qubit emission. The qubit is initialized by applying a resonant microwave drive with a fifth-order super-Gaussian envelope, see Fig.\ref{fig:emission_time_domain}b.  Formally, this drive pulse can be written as a classical field:
\begin{equation}
    S_{drive}\left(t\right)=De^{ \left( \frac{2t-\tau}{\tau} \right)^{2n}\ln{\alpha}}\sin(\omega_{01}t+\varphi_{drive})
\end{equation}
where $\tau$ is the pulse duration, $D$ is the pulse amplitude, $\omega_{01}$ is the qubit transition frequency, $\varphi_{drive}$ is the pulse phase, $n$ is the super-Gaussian order and $\alpha$ defines the amplitude of the pulse at its edges, allowing the super-Gaussian envelope to smoothly approach a small but finite value rather than zero. This is consistent with typical waveform generation constraints.  

The qubit is initially in the ground state  $\lvert g\rangle$, and at the end of the drive pulse the system state can be presented in the form of Eq. (\ref{eq:superpos_general}), as follows:
\begin{equation}
\label{eq:tehcnical_final_state}
\lvert\Psi(\tau)\rangle=\left(\cos\left(\frac{\theta}{2}\right)  \lvert g\rangle + \sin\left(\frac{\theta}{2}\right) e^{-i\varphi_{drive}}\lvert e\rangle\right)\otimes\lvert 0\rangle,
\end{equation}
where $\theta=\int_0^{\tau} {D e^{\left(\frac{2t-\tau}{\tau} \right)^{2n}\ln{\alpha}}dt}$ is the effective area of the drive pulse envelope. 
Following state preparation with a $16$~ns-long drive pulse, the emitted field is statistically monitored using time-resolved heterodyne detection. The measurement window spans 180~ns, including 20~ns prior to the drive pulse.  Each repetition $n$ yields the $I$ and $Q$ quadratures of the outgoing microwave field, $S_n(t) = I_n(t) + i Q_n(t)$, from which the complex temporal envelope of the emission is reconstructed. The repetitions are separated by $4$~$\mu$s, ensuring statistical independence between shots (see Fig.~\ref{fig:emission_time_domain}b for the pulse sequence).

To extract the single-shot phase information, each recorded trace $S_n(t)$ was digitally integrated over a fixed time window $T$, starting at $t_{\mathrm{start}}$:
\begin{equation}
\int_{t_{\mathrm{start}}}^{t_{\mathrm{start}}+T} S_n(t)\, dt
= \tilde{I}_n + i\tilde{Q}_n .
\end{equation}
The corresponding phase sample was calculated as
\begin{equation}
\varphi_n = \arg(\tilde{I}_n + i\tilde{Q}_n).
\end{equation}

Since $\varphi$ is a continuous variable defined on the interval 
$(0,2\pi]$, its statistics are characterized by the probability 
density function $\mathcal{P}(\varphi)$. In practice, 
$\mathcal{P}(\varphi)$ is obtained as the normalized histogram of the 
ensemble $\{\varphi_n\}_{n=1}^N$, such that
\begin{equation}
\int_{0}^{2\pi} \mathcal{P}(\varphi)\, d\varphi = 1 .
\end{equation}
For a finite ensemble, $\mathcal{P}(\varphi)$ is evaluated using a discrete binning with bin width $\Delta\varphi = 0.01\pi$, and normalized such that the sum over all bins equals unity.

\end{document}